\documentstyle[12pt]{article}

\newcommand{\EQ}{\begin{equation}}
\newcommand{\EN}{\end{equation}}
\newcommand{\bea}{\begin{eqnarray}}
\newcommand{\ena}{\end{eqnarray}}
\newcommand{\vs}[1]{\vspace{#1 mm}}
\newcommand{\hs}[1]{\hspace{#1 mm}}

\renewcommand{\d}{\delta}
\newcommand{\e}{\epsilon}
\def\bbox{{\,\lower0.9pt\vbox{\hrule \hbox{\vrule height 0.2 cm
\hskip 0.2 cm \vrule height 0.2 cm}\hrule}\,}}
\newcommand{\dsl}{\pa \kern-0.5em /}
\newcommand{\Dsl}{D \kern-0.6em /}

\newcommand{\shalf}{\frac{1}{2}}
\newcommand{\pa}{\partial}
\newcommand{\nn}{\nonumber\\}
\newcommand{\p}[1]{(\ref{#1})}
\newcommand{\lan}{\langle}
\newcommand{\ran}{\rangle}

\begin{document}
\topmargin 0pt
\oddsidemargin 0mm

\renewcommand{\thefootnote}{\fnsymbol{footnote}}
\begin{titlepage}

\setcounter{page}{0}
\begin{flushright}
OU-HET 308 \\
hep-th/9811057
\end{flushright}

\vs{10}
\begin{center}
{\Large\bf Thermalization of Poincar\'{e} Vacuum State and Fermion Emission
from AdS$_3$ Black Holes in Bulk-Boundary Correspondence}
\vs{10}

{\large
Nobuyoshi Ohta\footnote{e-mail address: ohta@phys.wani.osaka-u.ac.jp}
and
Jian-Ge Zhou\footnote{e-mail address: jgzhou@phys.wani.osaka-u.ac.jp,
JSPS postdoctoral fellow}}

\vs{10}
{\em Department of Physics, Osaka University,
Toyonaka, Osaka 560-0043, Japan}
\end{center}

\vs{10}
\centerline{{\bf{Abstract}}}
\vs{5}

The greybody factors for spin $\frac{1}{2}$ particles in the BTZ black holes
are discussed from 2D CFT in bulk-boundary correspondence. It is found that
the initial state of spin $\frac{1}{2}$ particle in the BTZ black holes can
be described by the Poincar\'{e} vacuum state in boundary 2D CFT, and the
nonlinear coordinate transformation causes the thermalization of the
Poincar\'{e} vacuum state. For special case, our results for the greybody
factors agree with the semiclassical calculation.

\end{titlepage}

\newpage
\renewcommand{\thefootnote}{\arabic{footnote}}
\setcounter{footnote}{0} 

In recent years there has been great progress in our understanding of black
hole physics from string and conformal field theories (CFT).
{} For reviews on this subject, see refs.~\cite{PC}. Through these studies,
it has been gradually realized that some 5D and 4D black holes contain BTZ
black holes~\cite{BTZ} in the near-horizon region~\cite{M,S1}, and
higher-dimensional black hole physics is essentially encoded in that of BTZ
black holes. In fact it has been shown in ref.~\cite{SSH} that the
Bekenstein-Hawking entropy for 5D and 4D black holes can be related to the
entropy of BTZ black holes by making use of U-duality.

Though much work on the absorption and Hawking radiation in 5D and 4D black
holes has also been made in the semiclassical analysis, D-brane picture,
effective string model and effective 2D CFT in refs.~\cite{DMW}-\cite{H},
only recently one began to recognize that not only the Bekenstein-Hawking
entropy but also the greybody factors in 5D and 4D black holes can be
understood as effectively coming from the near-horizon BTZ
geometry~\cite{T,MOZ,SA}. Since the recently discovered AdS/CFT
correspondence~\cite{M,GKP,W} might play an important role in the
fundamental quantum theory for the black holes via the near-horizon BTZ
black holes, it is natural to expect that the greybody factors in the black
holes should be elucidated in the context of the AdS/CFT correspondence.

Indeed the greybody factors for scalar fields have been derived by using
the near-horizon BTZ (AdS$_3$) geometry and AdS/CFT
correspondence~\cite{T,MOZ}. In ref.~\cite{MOZ}, the initial
state of scalar particles in BTZ black holes has
been described by a Poincar\'{e} vacuum state in the boundary 2D CFT.
The nonlinear coordinate transformation between the Poincar\'{e} coordinates
$(w_+,w_-)$ and the BTZ coordinates $(u_+,u_-)$ induces a mapping of the
operator ${\cal O}(w_+,w_-)$ to ${\cal O} (u_+,u_-)$ by Bogoliubov
transformation, and the operator ${\cal O}(u_+,u_-)$ sees the Poincar\'{e}
vacuum state as an excited mixed state, that is, they see the Poincar\'{e}
vacuum state as thermal bath of excitations in BTZ modes~\cite{MOZ,MS2,K}.
The usual procedure to thermally average the initial state of scalar
particles in the calculation of greybody factors is just to measure
the Poincar\'{e} vacuum state by the operator ${\cal O}(u_+,u_-)$ in the
BTZ coordinates.

In this paper, we discuss the greybody factors for spin $\shalf$ particle
in the BTZ black holes from AdS/CFT correspondence. Though there have
been many studies of correlation functions in the boundary theory from
the bulk-boundary correspondence~\cite{HSM,MVL,FMM}, these calculations
have mostly been performed in the Poincar\'{e} coordinates. Only a few
papers discussed the two-point correlation functions for scalar fields
in the BTZ coordinates~\cite{MOZ,K}.

To describe fermion emission from the BTZ
black holes in the spirit of AdS/CFT correspondence, we need to calculate
two-point correlation functions for spinor fields in the BTZ coordinates
(including all coefficients in the calculation). As we know, the bulk-boundary
Green functions for scalar fields in the BTZ coordinates only depend on the
differences of the coordinates $(\Delta u_+,\Delta u_-)$~\cite{MOZ,K},
and so it is manifestly invariant under the translations in the boundary
BTZ coordinates. However, for spinor case, though the bulk-boundary Green
function in Poincar\'{e} coordinates $(w_+,w_-)$ is a function of
$\Delta w_+$ and $\Delta w_-$, its form in the BTZ coordinates
$(u_+,u_-)$ not only depends on $\Delta u_+$ and $\Delta u_-$, but also
on $(u_+,u_-;u'_+,u'_-)$. The reason is that when we construct
bulk-boundary Green functions in the Poincar\'{e} coordinates, we apply an
element of the $O(3,1)$ isometry group of AdS$_3$ to move the singularity
from $y=\infty$ to an arbitrary point on the boundary which has to be
accompanied by a compensating local Lorentz transformation for spinor fields
to preserve the gauge fixing on the dreibein~\cite{HSM}, and this local
Lorentz transformation in the bulk breaks the manifest invariance of the
bulk-boundary Green functions under the translations in the boundary
BTZ coordinates. Due to this special behavior of the bulk-boundary
Green functions in the BTZ coordinates, it is highly nontrivial to check
whether the two-point correlation functions for spinor fields in the BTZ
coordinates take the expected form with translational invariance~\cite{G,MOZ}.
Remarkably we find that this is indeed true and that the greybody factors
for spinor fields in the BTZ black holes obtained
from AdS/CFT correspondence agree with the known results for a special case
including the coefficients.

Let us first consider two-point correlation functions of the operators
coupling to the boundary values of spinor fields in the Poincar\'{e}
coordinates. The AdS$_3$ metric in the Poincar\'{e} coordinates is
\bea
ds^2 = \frac{l^2}{y^2} ( dy^2 + dw_+ dw_-).
\label{poi}
\ena
For simplicity, we choose the radius of our AdS$_3$ space $l=1$ in the
following discussions.

\newcommand{\G}{\Gamma}
\newcommand{\s}{\sigma}
The free spinor action on AdS$_3$ is\footnote{
We take $\G^y = \s^3,\G^1 = \s^1,\G^2 = \s^2$ and $w^\pm=x^1 \mp i x^2$
in Euclidean case, $\G^0 = - i\G^2, w^\pm=x^1 \pm t$ in Minkowski case,
and $\G^\pm= (\G^1 \pm \G^0)/2$. Note that $dtdx= \shalf dw_+ dw_-$.}
\bea
S_0 = \shalf \int dy dw_+ dw_- \sqrt{g} {\bar \Psi}(\Dsl - m) \Psi.
\label{free}
\ena
For spinor fields, the action~\p{free} vanishes for the field configuration
that satisfies the equation of motion. The free action~\p{free} should be
supplemented by a boundary term~\cite{HSM}, which can be induced from
the Hamiltonian version of AdS/CFT correspondence~\cite{AF}
\bea
S_1 = \lim_{y\to 0} \frac{1}{4} \int dw_+ dw_- \sqrt{g_0} {\bar \Psi} \Psi,
\label{sur}
\ena
where $g_0$ is the determinant of the induced metric $y^{-4}$~\cite{HSM}.
It has been shown that the theory thus defined is equivalent to CFT on
the two-dimensional boundary even though the bulk action in AdS$_3$ looks
not conformally invariant in three dimensions~\cite{HSM}.

The bulk field $\Psi(y,w_+,w_-)$ and ${\bar \Psi}(y,w_+,w_-)$ can be
obtained from the boundary value $\psi(w_+,w_-)$ and ${\bar \psi}(w_+,w_-)$
by~\cite{HSM}
\bea
\Psi(y,w_+,w_-) &=& \frac{m+\shalf}{2 \pi} \int dw'_+ dw'_-
\left[ y \G^y + (w_+ - w'_+) \G^- + (w_- - w'_-) \G^+ \right] \nn
&\times& [ y^2 + (w_+ - w'_+)(w_- - w'_-)]^{-3/2 + m \G^y} y^{1-m\G^y}
 \psi(w'_+, w'_-), \nn
{\bar \Psi}(y,w_+,w_-) &=& \frac{m+\shalf}{2 \pi} \int dw'_+ dw'_-
 {\bar \psi}(w'_+, w'_-) y^{1+m\G^y} [ y^2 + (w_+ - w'_+)
(w_- - w'_-)]^{-3/2 - m \G^y} \nn
&\times& \left[ y \G^y + (w_+ - w'_+) \G^- + (w_- - w'_-) \G^+ \right].
\label{sol}
\ena

If we take mass $m$ positive, we have to choose $\G^y \psi(w_+,w_-) =
- \psi(w_+,w_-)$ and ${\bar \psi}(w_+,w_-) \G^y = {\bar \psi}(w_+,w_-)$
for the consistency of the theory~\cite{HSM}, which means that
$\psi(w_+,w_-)$ can be written as
\bea
\psi(w_+,w_-) = \left( \begin{array}{c}
0 \\
\psi_0 (w_+,w_-)
\end{array}
\right).
\label{com}
\ena
Then eq.~\p{sol} is recast into
\bea
\Psi(y,w_+,w_-) &=& \int dw_+' dw_-' K_P(y,w_+,w_-;w_+',w_-')
 \psi_0(w_+',w_-'), \nn
{\bar \Psi}(y,w_+,w_-) &=& \int dw_+' dw_-' \psi_0^\dagger (w_+',w_-')
 {\tilde K}_P(y,w_+,w_-;w_+',w_-'),
\label{sol1}
\ena
with the normalized bulk-boundary Green functions in the Poincar\'{e}
coordinates given by
\bea
K_P (y,w_+ - w_+',w_- - w_-') &=& \frac{m + \shalf}{2\pi} y^{m+1}
\left[ y^2 + (w_+ -w_+')(w_- -w_-') \right]^{-m-3/2} \nn
&& \hs{10} \times \left(
\begin{array}{c}
w_+ - w_+' \\
-y
\end{array}
\right), \nn
{\tilde K}_P (y,w_+ - w_+',w_- - w_-') &=& - \frac{m + \shalf}{2\pi}
 y^{m+1} \left[ y^2 + (w_+ -w_+')(w_- -w_-') \right]^{-m-3/2} \nn
&& \hs{10} \times \; ( y, w_+ - w_+' ),
\label{gf}
\ena
where $K_P$ and ${\tilde K}_P$ are manifestly invariant under the translations
in the boundary Poincar\'{e} coordinates $(w_+,w_-)$.

The coupling between the operators and the boundary
values of spinor fields takes the form
\EQ
S({\bar \psi},\psi) = \shalf \int dw_+ dw_- ( {\bar{\cal O}}\psi
 + {\bar \psi}{\cal O}) (w_+,w_-).
\label{cou}
\EN
Owing to eq.~\p{com}, without any loss of generality, we can choose
\bea
{\cal O}(w_+,w_-) = \left( \begin{array}{c}
{\cal O}_0 (w_+,w_-) \\
0
\end{array}
\right).
\label{com1}
\ena
The coupling~\p{cou} is then reduced to
\EQ
S = \shalf \int dw_+ dw_- ( - {\cal O}_0^\dagger \psi_0
 + \psi_0^\dagger {\cal O}_0 ) (w_+,w_-).
\label{cou1}
\EN
According to AdS/CFT correspondence, the relation between string theory
in the bulk and field theory on the boundary is~\cite{W}
\EQ
e^{-S_1({\bar \Psi}, \Psi)} = \lan e^{S({\bar \psi}, \psi)} \ran_{CFT}.
\label{for}
\EN
The two-point correlation function for spinor fields in the Poincar\'{e}
coordinates is~\cite{HSM}
\bea
\lan {\cal O}_0^\dagger (w_+,w_-){\cal O}_0(w_+',w_-')\ran =
 \frac{m + \shalf}{\pi} \frac{1}{(w_+ - w_+')^{2h_+}(w_- - w_-')^{2h_-}},
\label{cor}
\ena
with
\bea
h_+ = h - \frac{1}{4},\;\;
h_- = h + \frac{1}{4},\;\;
h = \frac{m+1}{2},\;\;
h_- - h_+ = \frac{1}{2},
\label{conf}
\ena
which shows that the conformal dimensions
for ${\cal O}_0^\dagger (w_+,w_-)$ and ${\cal O}_0 (w_+,w_-)$ are
$(h_+, h_-)$ when we take $m$ to be positive.

{}From the conformal invariance of the action~\p{cou1}, we know that
the conformal dimensions for $\psi_0^\dagger(w_+,w_-)$ and
$\psi_0(w_+,w_-)$ are $(1-h_+, 1-h_-)$.

Now we turn to the two-point correlation functions for spinor fields in the
BTZ coordinates. The metric of the BTZ black holes is~\cite{BTZ}
\bea
ds^2 = - \frac{(r^2-r_+^2)(r^2-r_-^2)}{r^2} dt^2
 + \frac{r^2}{(r^2-r_+^2)(r^2-r_-^2)} dr^2 + r^2 \left( d\phi
 - \frac{r_+ r_-}{r^2} dt \right)^2,
\label{met}
\ena
with periodic identification $\phi \sim \phi + 2 \pi$, where we have chosen
$l=1$. The mass and angular momentum are defined as
\bea
M = r_+^2 + r_-^2, \;\;
J = 2 r_+ r_-.
\ena

It has been shown that the metric of BTZ black holes can be transformed
to that of AdS$_3$ locally by the transformation which in the region 
$r>>r_\pm$ takes the form~\cite{MS2,K}
\bea
\label{coo1}
w_\pm &=& e^{2\pi T_\pm u_\pm}, \\
y &=& \left( \frac{r_+^2 - r_-^2}{r^2} \right)^\shalf
 e^{\pi ( T_+ u_+ + T_- u_-)},
\label{coo2}
\ena
with
\EQ
T_\pm = \frac{r_+ \mp r_-}{2\pi}, \;\;
u_\pm = \phi \pm t.
\label{coo3}
\EN

The boundary fields $\psi_0^\dagger (w_+,w_-)$ and $\psi_0(w_+,w_-)$
have conformal dimensions $(1-h_+,1-h_-)$, and so they transform as
\bea
\psi_0 (w_+,w_-) &=& \left( \frac{dw_+}{du_+}\right)^{h_+ -1}
 \left( \frac{dw_-}{du_-}\right)^{h_- -1} \psi_0(u_+,u_-), \nn
\psi_0^\dagger (w_+,w_-) &=& \left( \frac{dw_+}{du_+}\right)^{h_+ -1}
 \left( \frac{dw_-}{du_-}\right)^{h_- -1} \psi_0^\dagger (u_+,u_-),
\label{tra}
\ena
under the transformation~\p{coo1} and \p{coo2}.

Combining eqs.~\p{sol1}, \p{gf}, \p{coo1}, \p{coo2} and \p{tra}, we find
that the relation between the bulk and boundary fields in the BTZ coordinates
is changed into
\bea
\Psi(r,u_+,u_-) &=& \int du_+' du_-' K_B(r,u_+,u_-;u_+',u_-')
 \psi_0(u_+',u_-'), \nn
{\bar \Psi}(r,u_+,u_-) &=& \int du_+' du_-' \psi_0^\dagger (u_+',u_-')
 {\tilde K}_B(r,u_+,u_-;u_+',u_-'),
\label{sol2}
\ena
with
\bea
K_B (r,u_+, u_+';u_-, u_-') &=& \frac{2h - \shalf}{2\pi} (2\pi T_+)^{h_+}
 (2\pi T_-)^{h_-} e^{\frac{\pi}{2}(T_+ u_+ - T_- u_-)}
 \left( \frac{r_+^2 - r_-^2}{r^2} \right)^h \nn
&& \hs{-20} \times \left[ \frac{r_+^2 - r_-^2}{r^2}
 e^{\pi (T_+ \Delta u_+ + T_- \Delta u_-)}
 + 4 \sinh (\pi T_+ \Delta u_+) \sinh( \pi T_- \Delta u_-)
 \right]^{-(2h+\shalf)} \nn
&& \times \left( \begin{array}{c}
2 \sinh (\pi T_+ \Delta u_+) \\
-\sqrt{\frac{r_+^2 - r_-^2}{r^2}} e^{\pi (-T_+ u_+' + T_- u_-)}
\end{array} \right), \nn
{\tilde K}_B (r,u_+, u_+';u_-, u_-') &=& \frac{(2h - \shalf)}{2\pi}
 (2\pi T_+)^{h_+} (2\pi T_-)^{h_-} e^{\frac{\pi}{2}(T_+ u_+ - T_- u_-)}
 \left( \frac{r_+^2 - r_-^2}{r^2} \right)^h \nn
&& \hs{-20} \times \left[ \frac{r_+^2 - r_-^2}{r^2}
 e^{\pi (T_+ \Delta u_+ + T_- \Delta u_-)}
 + 4 \sinh (\pi T_+ \Delta u_+) \sinh( \pi T_- \Delta u_-)
 \right]^{-(2h+\shalf)} \nn
&& \times \left(
 - \sqrt{\frac{r_+^2 - r_-^2}{r^2}} e^{\pi (-T_+ u_+' + T_- u_-)},\ 
 - 2 \sinh (\pi T_+ \Delta u_+) \right),
\label{gf1}
\ena
and
\EQ
\Delta u_+ = u_+ - u_+', \;\;
\Delta u_- = u_- - u_-'.
\EN
Note that in contrast to $K_P(y, \Delta w_+, \Delta w_-)$ and ${\tilde K}_P
(y, \Delta w_+, \Delta w_-)$, $K_B$ and ${\tilde K}_B$ are not manifestly
invariant under the translations in the boundary coordinates $(u_+,u_-)$.
This is because when we apply an element of the $O(3,1)$ isometry group
of AdS$_3$ to construct the solution \p{sol}, we have to compensate
local Lorentz transformation for spinor fields to preserve the gauge
choice~\cite{HSM}, and this local Lorentz transformation in the bulk
breaks the manifest invariance of the bulk-boundary Green functions
under translations in the boundary BTZ coordinates. Nevertheless, we will
show that these Green functions give translationally invariant two-point
correlation functions in the boundary coordinates.

In the BTZ coordinates, the boundary action $S_1$ in \p{sur} turns into
\bea
S_1= \lim_{r\to \infty} \frac{1}{4} \int du_+ du_-
 \left( \frac{r^2}{r_+^2 - r_-^2} \right) {\bar \Psi}(r,u_+,u_-)
 \Psi(r,u_+,u_-).
\label{act1}
\ena

\newpage\noindent
Inserting eqs.~\p{sol2} and \p{gf1} into \p{act1}, we have\footnote{
In deriving \p{act2}, we have exploited the formula
\begin{eqnarray*}
&& \lim_{r\to \infty} \left( \frac{r_+^2 - r_-^2}{r^2}
 \right)^{2h-\shalf} \left[ \frac{r_+^2 - r_-^2}{r^2}
 e^{\pi (T_+ \Delta u_+ + T_- \Delta u_-)}
 + 4 \sinh (\pi T_+ \Delta u_+) \sinh( \pi T_- \Delta u_-)
 \right]^{-(2h+\shalf)} \\
&& \hs{20} = \frac{\pi}{ 2 h_+ (2\pi T_+)(2\pi T_-)}
 \d(\Delta u_+)\d(\Delta u_-).
\end{eqnarray*}
}
\bea
S_1 &=& \frac{h_+}{8\pi} \int du_+ du_- du_+' du_-'
 \psi_0^\dagger (u_+',u_-') \left( \frac{\pi T_+}{\sinh (\pi T_+ \Delta u_+)}
 \right)^{2h_+} \left( \frac{\pi T_-}{\sinh (\pi T_- \Delta u_-)}
 \right)^{2h_-} \nn
&& \hs{10} \times \; \psi_0(u_+,u_-).
\label{act2}
\ena
In the BTZ coordinates, the boundary action~\p{cou1} for the operators
${\cal O}_0^\dagger (w_+,w_-), {\cal O}_0 (w_+,w_-)$ and the boundary
values of spinor fields can be written as
\EQ
S = \shalf \int du_+ du_- ( - {\cal O}_0^\dagger \psi_0
 + \psi_0^\dagger {\cal O}_0 ) (u_+,u_-).
\label{cou2}
\EN
By using eq.~\p{for}, one finds
\bea
G(t,\phi) &=& \lan {\cal O}_0^\dagger(u_+,u_-) {\cal O}_0 (0,0) \ran \nn
&=&  \frac{h_+}{2\pi}
\left( \frac{\pi T_+}{\sinh \pi T_+ u_+} \right)^{2h_+}
\left( \frac{\pi T_-}{\sinh \pi T_- u_-} \right)^{2h_-},
\label{g1}
\ena
which indicates that the operators ${\cal O}_0^\dagger (u_+,u_-)$
and ${\cal O}_0 (u_+,u_-)$ see the Poincar\'{e} vacuum state as
an excited mixed state, that is, they see the Poincar\'{e} vacuum state
as thermal bath of excitations in BTZ modes~\cite{MOZ,MS2,K}.

Before proceeding, it would be appropriate to compare the above method to
derive \p{g1} with Unruh's calculation~\cite{UN}. Here the equation of motion
in the background of three-dimensional AdS space is solved and the bulk field
is expressed by the boundary value of the corresponding field with the help
of the normalized bulk-boundary Green function. By using the normalized Green
function in the BTZ coordinates, the conformal dimensions for boundary fields
and eqs.~\p{coo1} and \p{coo2}, we can obtain eq.~\p{g1}.
In Unruh's calculation for the fermion~\cite{UN}, on the other hand, the equation
of motion in two-dimensional collapsing-shell metric was solved and a conformal
transformation similar to \p{coo1} was exploited but not \p{coo2}, since
the equation of motion was analysed only in two dimensions. Thus the above
method in the context of AdS/CFT correspondence is essentially different from
Unruh's calculation, but relies on Witten's conjecture~\cite{W}.

It is known that the normalization factor cannot be fixed in the effective
CFT without recourse to string theory~\cite{G}. However with the help of
the normalized bulk-boundary Green function we can determine the normalization
factor for spin $\shalf$ particles. Also from the bulk-boundary correspondence,
the dependence of conformal dimensions on the AdS$_3$ mass $m$ can be read
off easily.

Because of the periodic identification of the coordinate $\phi$,
the two-point correlation function $G(t,\phi)$ should be modified
as~\cite{MOZ,K}
\bea
G_T(t,\phi) &=& \lan {\cal O}_0^\dagger (u_+,u_-) {\cal O}_0(0,0) \ran, \nn
&& \hs{-15} = \frac{h_+}{2\pi}
\sum_{n=-\infty}^{\infty}
\left( \frac{\pi T_+}{\sinh \pi T_+(\phi + t + 2n\pi) } \right)^{2h_+}
\left( \frac{\pi T_-}{\sinh \pi T_-(\phi - t + 2n\pi) } \right)^{2h_-},
\label{z1}
\ena
where the terms for $n \neq 0$ come from the twisted sectors of the
operators ${\cal O}_0^\dagger (u_+,u_-)$ and ${\cal O}_0(u_+,u_-)$ in
the orbifold procedure $u_\pm \sim u_\pm + 2n\pi$ for the BTZ black
holes~\cite{MS2}.

The greybody factors for spinor fields in the BTZ black holes are given
by~\cite{G,T}
\bea
\sigma_{abs} &=& \frac{2\pi}{\cal F} \int dt \int_{0}^{2\pi}d\phi
 e^{ip\cdot x} [ G_T(t-i\e,\phi) - G_T(t+i\e, \phi)] \nn
&=& \frac{2\pi}{\cal F} \int dt \int_{-\infty}^{\infty}d\phi
 e^{ip\cdot x} [ G(t-i\e,\phi) - G(t+i\e, \phi)] \nn
&=& \frac{h_+}{\cal F} \frac{(2\pi T_+)^{2h_+-1}(2\pi T_-)^{2h_- -1}}
{\Gamma(2h_+) \Gamma(2h_-)} \cosh\left( \frac{\omega}{2 T_H} \right) \nn
&& \times \left| \Gamma\left(h_+ + i\frac{\omega}{4\pi T_+} \right)
 \Gamma\left(h_- + i\frac{\omega}{4\pi T_-} \right) \right|^2,
\label{genf}
\ena
where ${\cal F}$ is the incident fermion flux and we have assumed the scaling
dimension for spinor fields is half an odd integer, which is true when we
consider the mass term $m$ induced from the Kaluza-Klein reduction from
AdS$_3 \times S^3\times M_4$~\cite{DKS}.
The Hawking temperature $T_H$ is defined by
\EQ
\frac{2}{T_H} = \frac{1}{T_+} + \frac{1}{T_-}.
\EN

The above calculation for
greybody factors has been performed in the 2D CFT on the two-dimensional
boundary, where the metric is given as $ds^2= dw_+ dw_-$ in the boundary
Poincar\'{e} coordinates $(w_+,w_-)$~\cite{HSM}. Thus we can use eq.~\p{genf}
to derive greybody factors from the point of view of 2D CFT.

Usually the greybody factors describe scattering of asymptotic states from
asymptotically flat black holes. Since AdS spacetimes do not have asymptotic
states in the same sense as in asymptotically flat spacetimes, we should
explain what ``greybody factors for BTZ black holes'' mean. In the limit of
large number $N$ of D-branes, the geometries of the 5D and 4D black holes
are BTZ $\times S^3 \times M_4$ and BTZ $\times S^2 \times M_5$,
respectively~\cite{MS2,T}, but they are asymptotically flat. For low-energy
emission, the greybody factors in higher-dimensional black holes computed
in gravity for asymptotically flat black holes are related to the two-point
correlation functions obtained in large $N$ D-brane gauge theories~\cite{DMW}.
In the context of AdS/CFT correspondence, these two-point functions can in
turn be computed from semiclassical gravity inside the throat region which
becomes BTZ black hole in a suitable limit. Thus we can do a classical gravity
calculation to compute the large $N$ gauge theory two-point correlation
functions which give the greybody factors for asymptotically flat black
holes. All of this shows that the greybody factors in higher-dimensional
black holes have their origin in BTZ black holes~\cite{T,MOZ,SA}.
The boundary dynamics of BTZ black holes, which is controlled by 2D CFT,
looks like hologram and contains the essential informations of
higher-dimensional black holes~\cite{MOZ}.

On the other hand, in asymptotically flat spacetime, the asymptotic observer
measures a decay rate which is modified by the greybody factor of the black
hole. To define the Hawking emission rate for the BTZ black hole in the gravity
calculations, we do not take an asymptotic observer, but an observer stationed
at $r \sim l >> r_+$, which means that we take the incoming flux in the region
$r \sim l >> r_+$ as the incident flux on the black hole~\cite{BSS,D}.
The reason for this choice is that in curved spacetime, an observer measures
a thermal spectrum depending on his/her local temperature $T_H/\sqrt{g_{00}}$,
and for asymptotically flat spacetime $\sqrt{g_{00}} \to 1$ as $r\to \infty$.
For BTZ black holes with $r_+ << l$, we see that $\sqrt{g_{00}} \to 1$ when
$r\sim l$, so the observer in BTZ geometry measures a local temperature equal
to the Hawking temperature at this position~\cite{D}.

To compare the above result with that from the semiclassical calculation,
we consider $(h_+,h_-) = (\shalf, 1)$ case. According to the definition
for ${\cal F}$ in refs.~\cite{G,D}, we can choose ${\cal F}=1$.
We find from eq.~\p{genf}
\bea
\sigma_{abs}^{(\shalf,1)} = \frac{\pi^2\omega}{4} \frac{\cosh\left(
 \frac{\omega}{2T_H} \right)}{\cosh\left( \frac{\omega}{4 T_+} \right)
 \sinh\left( \frac{\omega}{4 T_-} \right)}.
\label{genf1}
\ena
The emission rate of the fermions is given by the product with thermal
distribution:
\bea
\G^{(\shalf,1)} = \frac{\pi^2 \omega}{2}
 \frac{d^2 k}{\left[\exp\left(\frac{\omega}{2T_+}\right) + 1 \right]
 \left[\exp\left(\frac{\omega}{2T_-}\right) - 1 \right]},
\ena
which agrees with the semiclassical gravity calculations in ref.~\cite{D}
in near-extremal limit $(T_- >> T_+)$ and for energies small in comparison
with the size of the black holes (see eq.~(34) there).\footnote{
In ref.~\cite{D}, this has been evaluated effectively for the mass $m=\shalf$
(see eq.~(9) there and note here that the parameter $l$ has been set to 1),
which corresponds precisely to $h_+=\shalf, h_-=1$ according to eq.~\p{conf}.}

The agreement of greybody factors for spinor fields in the BTZ black holes
obtained from AdS/CFT correspondence with that from semiclassical gravity
calculations~\cite{D} supports the identification that the initial state
of the particle in the BTZ black holes can be described by the Poincar\'{e}
vacuum state in the boundary 2D CFT~\cite{MOZ}, and the nonlinear coordinate
transformation~\p{coo1} and \p{coo2} causes the thermalization of the
Poincar\'{e} vacuum state which holds valid also for the spinor case.
We believe that such an identification should also work for spin $\frac{3}{2}$
Rarita-Schwinger fields in BTZ black holes. However, in the above discussions,
we have only considered free fields. It would be interesting to check whether
such an identification works for interacting theories.

In order to compare the fermion emission from black holes with three charges
$(Q_1,Q_5,n)$ in five-dimensional $N=8$ supergravity between the above
AdS/CFT approach and semiclassical analysis~\cite{H}, we need to calculate
two-point correlation functions for spin $\shalf$ particle in the
near-horizon geometry of 5D black holes. By analogy with the scalar case, we
expect that it has the form~\cite{MOZ,FMM}
\bea
\lan {\cal O}_0^\dagger (u_+,u_-){\cal O}_0(0,0)\ran =
 \eta_{5D} \frac{h_+}{2\pi}
\left( \frac{\pi T_+}{\sinh\pi T_+ u_+}\right)^{2h_+}
\left( \frac{\pi T_-}{\sinh\pi T_- u_-}\right)^{2h_-}.
\label{cor1}
\ena
It is nice to see how to determine the coefficient $\eta_{5D}$ for
the spinor field precisely.

A recent work~\cite{ST} suggested that in the very near horizon limit
the above boundary 2D CFT is transformed into discrete light-cone
quantization of a CFT which has a connection with matrix model~\cite{BFSS}.
It would be interesting to see how the above fermion emission from BTZ
black holes in AdS/CFT correspondence is related to that in the context
of matrix black holes~\cite{BFKS}. Work along this line is under
investigation.

\section*{Acknowledgements}

We would like to thank Y. Satoh for useful discussions. This work was
supported in part by the grant-in-aid from the Ministry of Education,
Science, Sports and Culture No. 96208.

%\newpage
\newcommand{\NP}[1]{Nucl.\ Phys.\ {\bf #1}}
\newcommand{\AP}[1]{Ann.\ Phys.\ {\bf #1}}
\newcommand{\PL}[1]{Phys.\ Lett.\ {\bf #1}}
\newcommand{\CQG}[1]{Class. Quant. Gravity {\bf #1}}
\newcommand{\NC}[1]{Nuovo Cimento {\bf #1}}
\newcommand{\CMP}[1]{Comm.\ Math.\ Phys.\ {\bf #1}}
\newcommand{\IJMP}[1]{Int.\ Jour.\ Mod.\ Phys.\ {\bf #1}}
\newcommand{\JHEP}[1]{J.\ High\ Energy\ Phys.\ {\bf #1}}
\newcommand{\PR}[1]{Phys.\ Rev.\ {\bf #1}}
\newcommand{\PRL}[1]{Phys.\ Rev.\ Lett.\ {\bf #1}}
\newcommand{\PRE}[1]{Phys.\ Rep.\ {\bf #1}}
\newcommand{\PTP}[1]{Prog.\ Theor.\ Phys.\ {\bf #1}}
\newcommand{\PTPS}[1]{Prog.\ Theor.\ Phys.\ Suppl.\ {\bf #1}}
\newcommand{\MPL}[1]{Mod.\ Phys.\ Lett.\ {\bf #1}}
\newcommand{\JP}[1]{Jour.\ Phys.\ {\bf #1}}

\end{document}